\documentclass{PoS}
\title{New Measurement of the $B^0_s$ Mixing Phase at CDF }

\ShortTitle{$B^0_s$ mixing phase at CDF}

\author{\speaker{Gavril Giurgiu}\thanks{for the CDF collaboration}\\
        Johns Hopkins University\\
        E-mail: \email{ggiurgiu@jhu.edu}}

\abstract{
	   The CDF collaboration presents an updated measurement of the CP-violating 
	parameter $\beta_s^{J/\psi\phi}$ and of the decay width difference $\Delta\Gamma_s$ using  
	approximately 6500 $B_s \rightarrow J/\psi \phi$ decays collected by the dimuon trigger 
	and reconstructed in a data sample corresponding to 5.2~fb$^{-1}$ of integrated luminosity. 
	Besides exploiting the two-fold increase in the data sample with respect to the previous 
	measurement, several improvements have been introduced in the analysis including a fully 
	data-driven flavor tagging calibration and proper treatment of possible S-wave contributions.

	   We find that the CP-violating phase is within the range  
	$\beta_s^{J/\psi\phi} \in [0.02, 0.52] \cup [1.08,1.55]$ at 68\%~C.L. 
	The decay width difference is found to be 
	$\Delta\Gamma_s = 0.075~\pm~0.035$~(stat)~$\pm 0.01$~(syst)~ps$^{-1}$. 
	In addition, we present the most precise mean $B_s$ lifetime $\tau_s$, polarization 
	amplitudes $|A_0|^2$, $|A_{\parallel}|^2$ and $|A_{\perp}|^2$, 
	as well as strong phase $\delta_{\perp}$:
	\begin{center}
		\item[] $\tau_s = 458.6 \pm 7.6$ (stat) $\pm 3.6$ (syst)$~\mu$m
		\item[] $|A_0|^2 = 0.524 \pm 0.013 $ (stat) $\pm 0.015$ (syst)
		\item[] $|A_{\parallel}|^2 = 0.231 \pm 0.014 $ (stat) $\pm 0.015$ (syst)
		\item[] $\delta_{\perp} = 2.95 \pm 0.64 $ (stat) $\pm 0.07$ (syst). 	
	\end{center}

	}

\FullConference{35th International Conference of High Energy Physics - ICHEP2010,\\
		July 22-28, 2010\\
		Paris, France}

\begin{document}

\section{Introduction}

Charge-Parity (CP) violation is accommodated within the Standard Model (SM) 
via complex phases that enter mixing matrices which connect different family 
fermions. In the lepton sector the Pontecorvo-Maki-Nakagawa-Sakata 
(PNMS) matrix connects leptons from different generations, while in the 
quark sector, the Cabibbo-Kobayashi-Maskawa (CKM) matrix~\cite{ckm_matrix} 
mixes quarks from different generations and introduces CP violation in the 
hadronic sector. 
 
While CP violation has been well measured and found to agree with the SM expectations 
in Kaon and in most B-meson decays, the study 
of CP violation in decays of $B_s$ mesons is still in its early stages, with the 
first results from $B_s \rightarrow J/\psi \phi$ decays reported by the CDF and D0 
collaborations in the last couple of years~\cite{cdf_beta_s_prl,d0_beta_s_prl}. 
In these decays, CP violation occurs through the interference between the decay amplitudes 
with and without mixing. In the SM the relative phase between the decay amplitudes with and 
without mixing is $\beta_s^{SM}=\arg(-V_{ts}V_{tb}^{*}/V_{cs}V_{cb}^{*})$ and it 
is expected to be very small~\cite{bigi-sanda,Ref:lenz}. New physics contributions 
manifested in the $B_s^0$ mixing amplitude may alter this mixing phase by a 
quantity $\phi_s^{NP}$ leading to an observed mixing phase 
$2\beta_s^{J/\psi\,\phi} = 2\beta_s^{SM} - \phi_s^{NP}$. 
Large values of the observed $\beta_s^{J/\psi\,\phi}$ would be an indication of physics 
beyond the SM~\cite{Ref:lenz,Ref:hou,Ref:ligeti,theory}.        
                   
The analysis presented in this paper is an update of previous CDF analyses with 
1.4~fb$^{-1}$~\cite{cdf_beta_s_prl} and 2.8~fb$^{-1}$~\cite{cdf_beta_s_28fb} of 
integrated luminosity. A combination~\cite{cdf_d0_combination_28fb} between 
the 2.8~fb$^{-1}$ CDF result~\cite{cdf_beta_s_28fb} and the corresponding one from D0~\cite{d0_beta_s_28fb} 
led to the conclusion that, assuming the standard model predictions~\cite{theory}, 
the probability of a deviation as large as the level of the observed data is 3.4\% 
corresponding to 2.1 Gaussian standard deviations. 
In that light, updated measurements from CDF and D0 were expected with more integrated 
luminosity. In this paper we present the 5.2~fb$^{-1}$ 
$\beta_s^{J/\psi\phi}$ results from CDF~\cite{cdf_beta_s_52fb}, first presented 
at FPCP 2010~\cite{louise_fpcp_2010}. 
The D0 update was shown at ICHEP 2010~\cite{d0_beta_s_52fb} and reveals 
an improvement of the $\beta_s^{J/\psi\phi}$ agreement with the SM expectation, at least 
as long as no other constraints are imposed. Adding external constraints, a 6-7.5\% 
agreement with SM is quoted.

\section{Measurement of CP violation in $B_s \rightarrow J/\psi\phi$ decays}

\begin{figure}[tb]
\includegraphics[width=75mm]{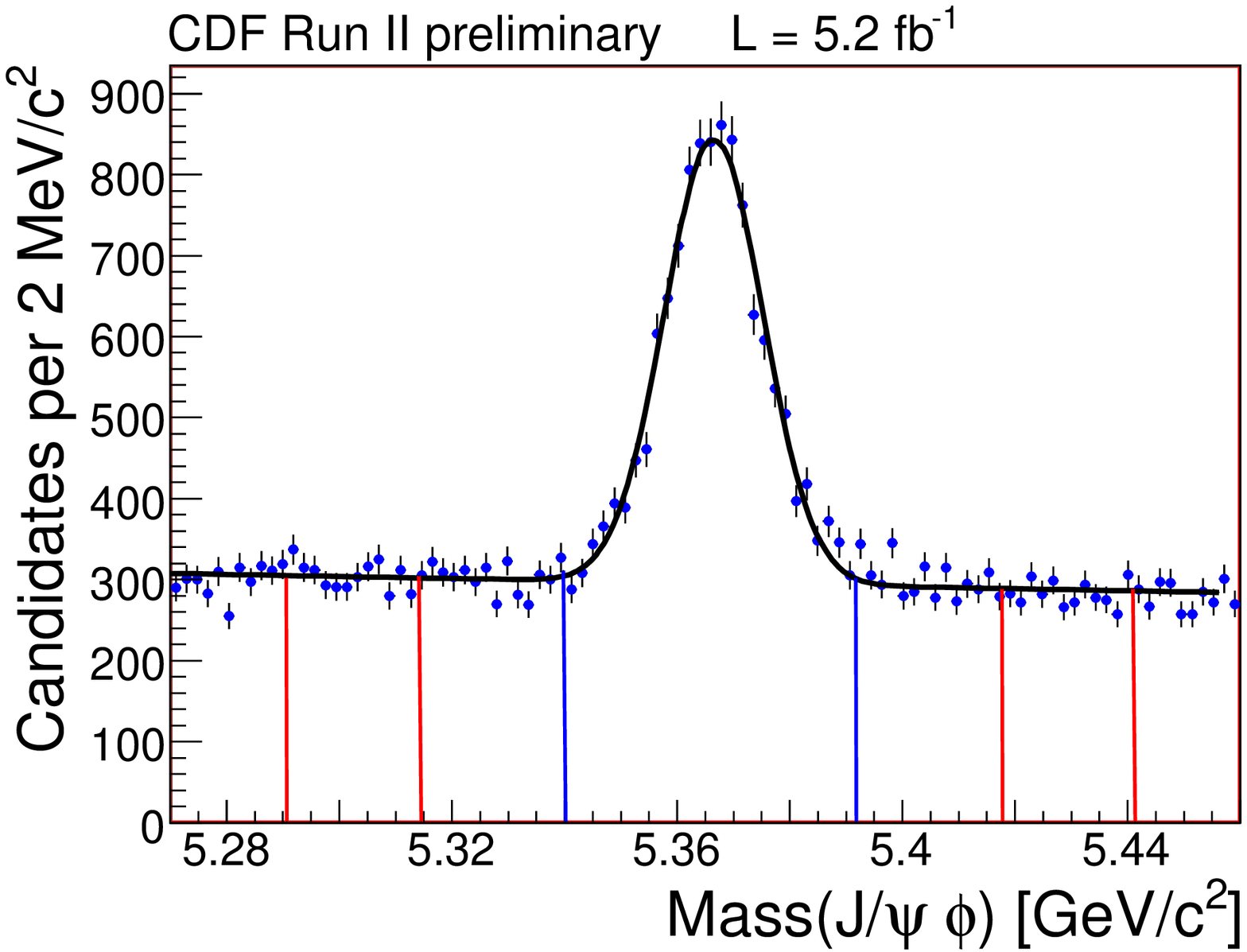}
\includegraphics[width=65mm]{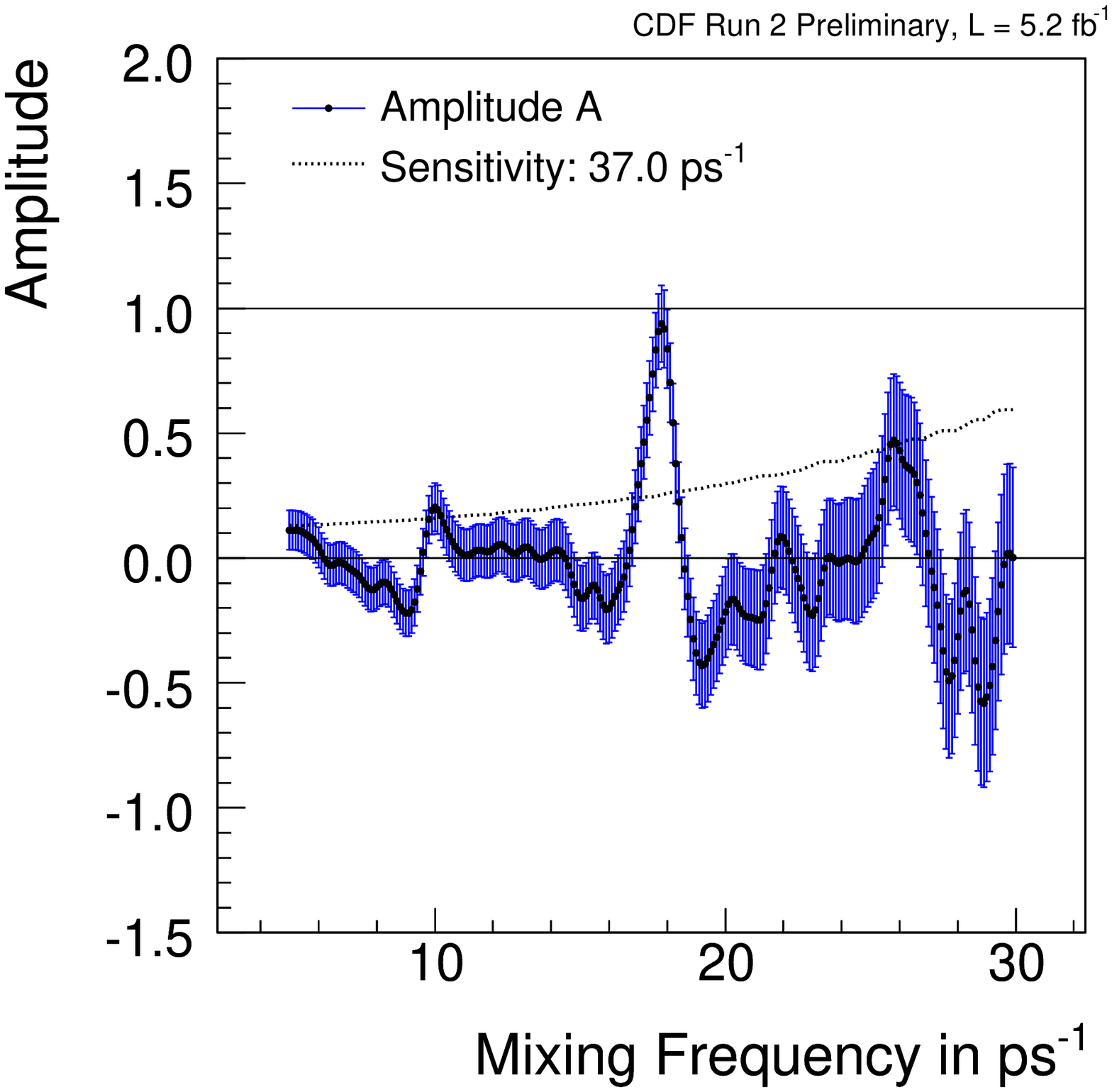}
\caption{ Left: $\mu^+ \mu^- K^+ K^-$ invariant mass distribution and fit projection. 
	Right: $B_s$ mixing amplitude scan with statistical errors only. }
\label{fig:Bs_mass}
\end{figure}

We reconstruct $B_s \rightarrow J/\psi\phi$ decays, 
where the $J/\psi$ meson decays in two opposite charge muons and the $\phi$ meson 
decays in two opposite charge kaons. Kinematic, topological and particle identification 
information (time of flight and energy loss) are combined into an artificial neural network 
optimized for $B_s \rightarrow J/\psi\phi$ candidate selection. The optimal 
selection is determined using pseudo-experiments in such a way that the expected 
statistical uncertainty on the $\beta_s^{J/\psi\phi}$ phase is minimized. We select a sample of    
$\approx 6500$ $B_s \rightarrow J/\psi\phi$ candidates with a signal to background 
ratio of $\approx 1$. Figure~\ref{fig:Bs_mass} shows the $\mu^+ \mu^- K^+ K^-$ invariant 
mass distribution along with the fit projection superimposed.

Having identified $B_s \rightarrow J/\psi\phi$ candidates, in order to improve the analysis 
sensitivity to CP violation parameter $\beta_s^{J/\psi\phi}$, we need to separate 
$B_s$ mesons produced as $B_s^0$ from the ones produced as $\bar{B_s^0}$.
We employ algorithms called flavor taggers 
which, on a candidate by candidate basis, provide the $B_s$ production flavor (tag) 
along with the expected probability $P$ that the tag is correct. Often, instead 
of the tagging probability, a derived quantity is quoted, the dilution: $D = 2P -1$.    
A tagging algorithm is also characterized by its efficiency $\epsilon$ which is the 
fraction of tagged $B_s$ mesons out of all $B_s$ candidates. The efficiency multiplied 
by the squared dilution provides the tagging effectiveness: a sample of $N$ events 
tagged by a tagging algorithm with efficiency $\epsilon$ and dilution $D$ has the 
same statistical power as a sample of $\epsilon D^2 N$ events identified with 100\% 
efficiency and 100\% tagging probability.

There are two types of flavor tagging algorithms 
used in this analysis. The first one takes advantage of the fact that in $p\bar{p}$ collisions 
at the Tevatron, $b$ and $\bar{b}$ quarks are dominantly produced in pairs. The opposite 
side tagging (OST) algorithm searches for leptons or jets opposite to  
the reconstructed $B_s$ meson. The opposite side is defined as the region outside 
the $\Delta R = \sqrt{\Delta\eta^2 + \Delta\phi^2}$ cone around the $B_s$ flight direction, 
where $\Delta\eta$ and $\Delta\phi$ define the pseudo-rapidity and the azimuthal direction 
with respect to $B_s$ direction. 
The OST algorithm uses either the lepton charge to infer the $B_s$ production flavor  
or the opposite side jet which contains a displaced vertex or track and uses 
the sum of the jet track charges, weighted by the track momenta, to identify the $B_s$ 
production flavor. The OST is calibrated on a sample of $\approx 52000$ $B^{\pm} \rightarrow J/\psi K^{\pm}$ 
decays~\cite{cdf_beta_s_52fb}. Here, the $B$ meson flavors at production and decay are the 
same. The flavor at decay is given by the kaon charge providing a straightforward way 
of measuring the 
tagging dilution. The ratio between the measured and predicted dilution is found to be
$1.02 \pm 0.07$ which is used as a calibrating factor in the $\beta_s^{J/\psi\phi}$ 
measurement. The OST efficiency is $(94.2 \pm 0.4)\%$ while the average dilution 
is $(11.5 \pm 0.2)\%$ for a total tagging power $\epsilon D^2 = 1.2\%$.   
Besides the OST we also used a flavor tagging method called same side tagging (SST). 
This algorithm searches for charged kaons produced during the hadronization process 
in which the $B_s$ meson is formed. The charge of the kaon is used to infer the 
$B_s$ flavor. The tagging dilution is determined from simulated events and calibrated 
in data by simultaneously measuring the $B_s$ mixing frequency and the SST dilution 
scale factor using $\approx 13000$ fully reconstructed $B_s \rightarrow D_s (3)\pi$ decays, 
where the $D_s$ meson decays either into $\phi \pi$ or into three pions~\cite{sskt_calibration}. 
Figure~\ref{fig:Bs_mass} shows the $B_s$ mixing amplitude as a function of the 
assumed oscillation frequency. The mixing amplitude is a coefficient that multiplies 
the mixing term $cos(\Delta m_s t)$ which describes the $B_s$ oscillations. Whenever the 
mixing frequency is fixed to a wrong value, the mixing amplitude is consistent with zero. 
When the oscillation frequency reaches the true value, the amplitude becomes inconsistent 
with zero. If the tagging algorithm provides accurate predicted dilution, then, the amplitude 
is consistent with one. Figure~\ref{fig:Bs_mass} shows that at 
$\Delta m_s = 17.79 \pm 0.07$ the amplitude is indeed consistent with unity: 
$A~=~0.94~\pm~0.15$~(stat)~$\pm 0.07$~(syst). This value is used as a dilution scale factor 
in the  $\beta_s^{J/\psi\phi}$ measurement
The SST efficiency is (52.2 +/- 0.7)\% and the average dilution is (27.5 +/- 0.3)\%. 
Accounting for the dilution calibration factor we find the total SST power to be $(3.5 \pm 1.4)\%$ 
It is worth noting that the measured oscillation frequency 
is consistent with the previous CDF~\cite{dms_cdf} and D0~\cite{dms_d0} mixing measurements.   

\begin{figure}[tb]
\includegraphics[width=75mm]{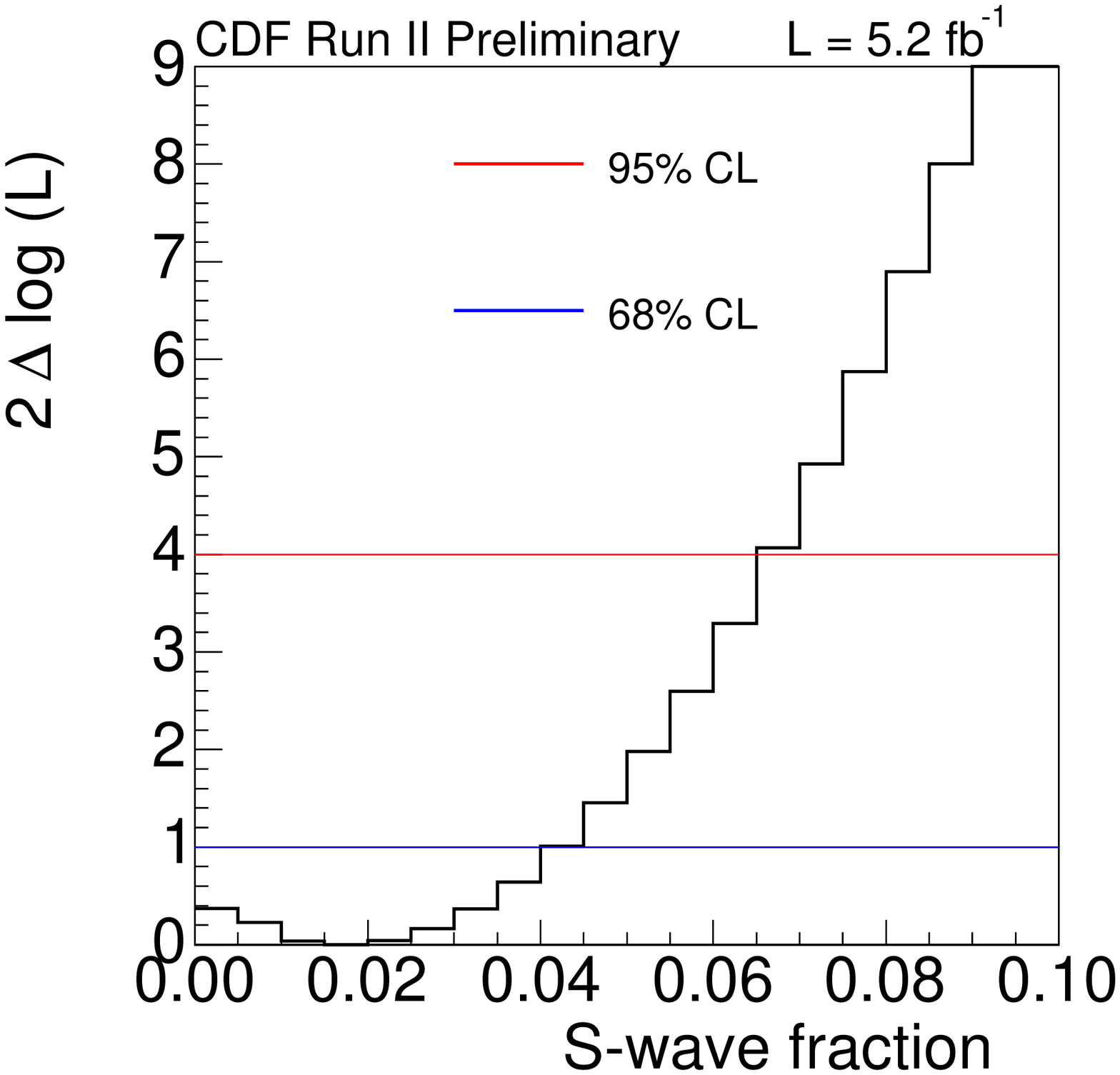}
\includegraphics[width=75mm]{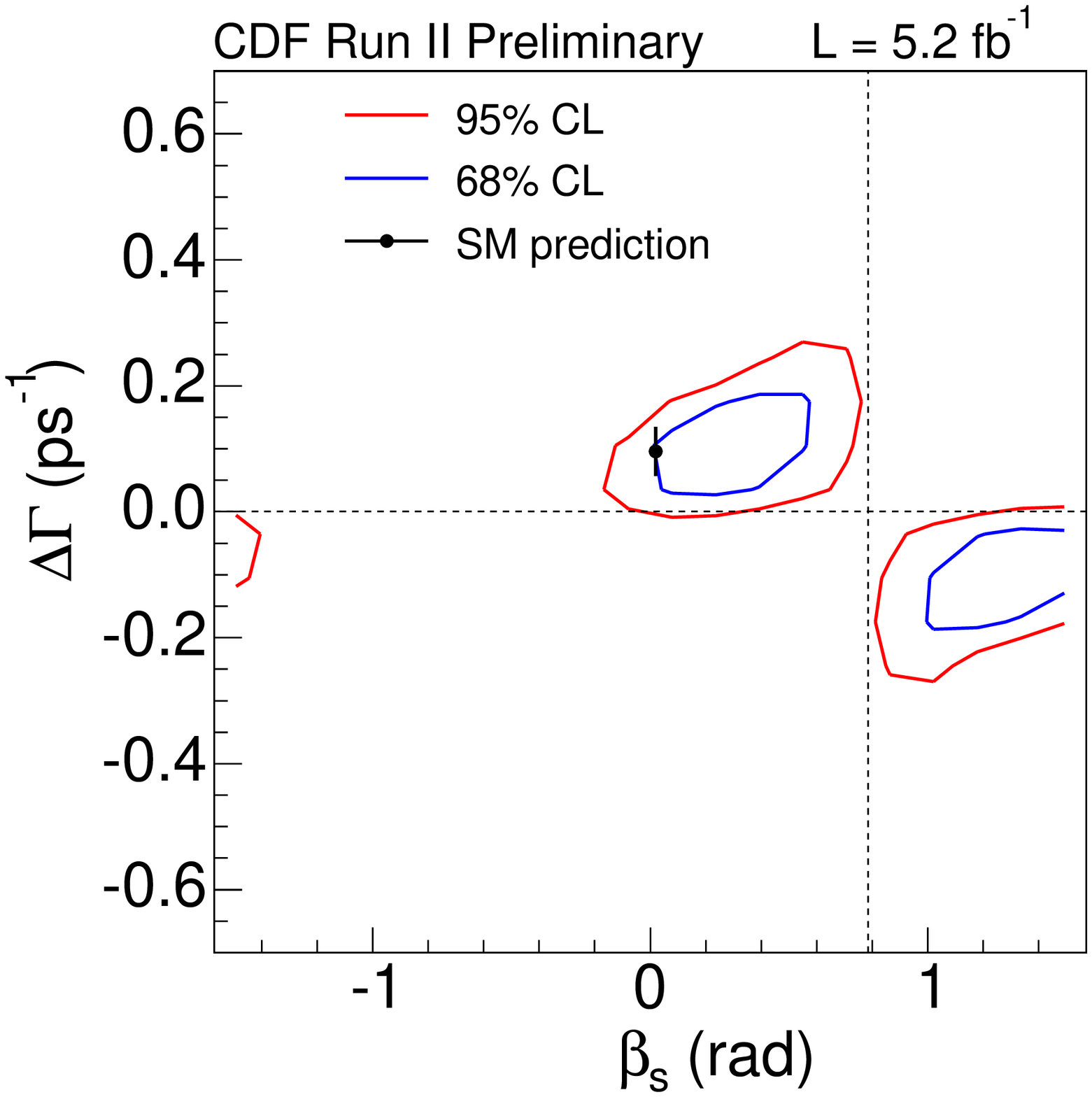}
\caption{ Left: Likelihood profile of the S-wave fraction.  
	Right: 68\% and 95\% confidence regions in the $\beta_s^{J/\psi\phi}-\Delta\Gamma$ plane. }
	\label{fig:beta_s}
\end{figure}

Another important component in this analysis is the statistical separation of CP event and CP odd $J/\psi\phi$ 
final states. While the $B_s$ meson has spin 0, the final states $J/\psi$ and $\phi$ 
have spin 1. Consequently, the total angular momentum in the final state 
can be either 0, 1 or 2. States with angular momentum 0 and 2 are CP 
even while the state with angular momentum 1 is CP odd. Angular distributions 
of the final muons and kaons from $J/\psi$ and $\phi$ decays can be used 
to statistically separate the CP eigenstates.    
There are three angles that completely define the directions of the 
four particles ($\mu^+, \mu^-, K^+, K^-$) in the final state. We use the angles  
$\vec\rho = \{\cos\theta_T,\phi_T, \cos\psi_T\}$
defined in the transversity basis introduced in Ref.~\cite{Ref:Dighe}.   
The detector acceptance is not flat in these angles. Corrections are introduced 
in the maximum likelihood fit to account for such effects. The correction 
method is validated by measuring the lifetime and polarization amplitudes in 
$B^0 \rightarrow J/\psi K^{*0}$ decays~\cite{cdf_d0_b0jpsikst}.

An unbinned maximum likelihood fit~\cite{jhep_paper} is performed to extract the
parameters of interest, $\beta_s^{J/\psi\phi}$ and $\Delta\Gamma_s$, 
plus additional parameters referred to as ``nuisance parameters'' which 
include the signal fraction 
$f_s$, the $B_s^0$ mass, the mean $B_s^0$ width $\Gamma\equiv (\Gamma_L + \Gamma_H)/2 = 1/\tau(B_s^0)$,
the magnitudes of the polarization
amplitudes in the transversity basis $|A_0|^2$, $|A_{\parallel}|^2$, and $|A_{\perp}|^2$, and
the strong phases $\delta_{\parallel}\equiv
\arg(A_{\parallel}A_0^{*})$ and $\delta_{\perp}\equiv
\arg(A_{\perp}A_0^{*})$. The fit uses information on the
reconstructed $B_s^0$ candidate mass,
the $B_s^0$ candidate proper decay time and its uncertainty, the transversity 
angles $\vec\rho$
and tagging information~\cite{dms_cdf} $P$ and $\xi$, where
$P$ is the event-by-event correct tag probability and $\xi=\{-1,0,+1\}$ is the
tag decision, in which $+1$ corresponds to a candidate tagged as
$B_s^0$, $-1$ to a $\bar{B}_s$, and $0$ to an untagged candidate.  

In this updated of the analysis we include in the maximum likelihood fit 
potential S-wave contributions from either $B_s \rightarrow J/\psi f_0$ with 
$f_0 \rightarrow K^+K^-$ or non-resonant $B_s \rightarrow J/\psi K^+K^-$ decays. 
While it has been suggested~\cite{stone_s_wave} that the S-wave contribution 
within $\pm 10$~MeV around the $\phi$ mass may be as large as 6.3\%, our analysis 
shows that the S-wave contribution is less that 6.7\% at 95\% confidence level. 
Figure~\ref{fig:beta_s} shows the likelihood profile of the S-wave fraction 
with a best fit value of $\approx 2\%$.

\section{Results and Conclusions}

Assuming no CP violation ($\beta_s^{J/\psi\phi} = 0$), the $B_s$ decay width difference, 
lifetime, polarization fractions and strong phase $\delta_{\perp}$ 
are measured and found to be the most precise to date:
$\Delta\Gamma_s = 0.075 \pm 0.035 $ (stat) $\pm 0.01$ (syst) ps$^{-1}$, 
$\tau_s = 458.6 \pm 7.6$ (stat) $\pm 3.6$ (syst)$~\mu$m, 
$|A_0|^2 = 0.524 \pm 0.013 $ (stat) $\pm 0.015$ (syst), 
$|A_{\parallel}|^2 = 0.231 \pm 0.014 $ (stat) $\pm 0.015$ (syst), 
$\delta_{\perp} = 2.95 \pm 0.64 $ (stat) $\pm 0.07$ (syst).
The strong phase $\delta_{\parallel}$ is close to the symmetry point at $\pi$ 
which makes the estimation of its value and uncertainty unreliable and therefore 
we do not provide a result for it at this point. 
Finally, the confidence regions in the $\beta_s^{J/\psi\phi}-\Delta\Gamma_s$ plane are 
presented in Figure~\ref{fig:beta_s}. 
The ambiguity between the two minima could be resolved if the strong phases
$\delta_{\parallel}$ and $\delta_{\perp}$ were known. Recent theoretical studies~\cite{Ref:gronau}
suggest that the strong phases involved in $B_s^0 \rightarrow J/\psi\,\phi$ 
decays are expected to be close to the corresponding strong phases in 
$B^0 \rightarrow J/\psi\,K^{*0}$. Using this information, the preferred solution would be 
the one corresponding to positive $\Delta\Gamma_s$. 
For the SM values of $\beta_s^{J/\psi\phi}$ and $\Delta\Gamma_s$ we obtain a p-value of 0.44 
which corresponds to about 0.8 standard deviations. 
Minimizing the likelihood function over $\Delta\Gamma_s$ as well, we find  
$\beta_s^{J/\psi\phi} \in ([0.02, 0.52] \cup [1.08, 1.55])$~at~68\%~C.L.
and $\beta_s^{J/\psi\phi} \in ([-\pi/2, -1.44] \cup [-0.13,0.68] \cup [0.89, \pi/2])$~at~95\%~C.L.
The p-value for the SM expectations of $\beta_s^{J/\psi\phi}$ is 0.31. 
\\

We have presented an updated measurement of the CP violation parameter $\beta_s^{J/\psi \phi}$ 
in $B_s \rightarrow J/\psi\phi$ decays using 5.2~fb$^{-1}$ of integrated luminosity accumulated 
by the CDF experiment. The updated result provides bounds on $\beta_s^{J/\psi \phi}$ which 
are significantly stronger than previous results. The data are consistent with the SM expectations 
at one standard deviation level. Due to multiple improvements to the analysis, the 
improvement compared to the 2.8~fb$^{-1}$ result is better than the expectation from just scaling 
the previous results by the amount of additional data sample. Future analyses may add other decay 
modes and data from other CDF triggers. The data sample is expected to double by 2011.

\end{document}